\documentclass{emulateapj}

\usepackage{epsf}
\usepackage{graphics}
\usepackage{color}

\newcommand{\be}{\begin{equation}}
\newcommand{\ee}{\end{equation}}

\newcommand{\bx}{$\beta_{\rm X}$}

\newcommand{\plm}{$\pm$}
\newcommand{\nh}{$N_{\rm H}$}
\newcommand{\swift}{\mbox{\it Swift}}	    
\newcommand{\chandra}{\mbox{\it Chandra}}	    
\newcommand{\meszaros}{M\'esz\'aros~}

\shorttitle{Chandra Observations of GRB050724}
\shortauthors{Grupe et al.}

\begin{document}


\def\etal{{\it et\thinspace al.}\ }
\def\alp{{$\alpha$}\ }
\def\al2{{$\alpha^2$}\ }

%
%
%


\title{Jet Breaks in Short Gamma-Ray Bursts.  I: The Uncollimated
  Afterglow of GRB\,050724}


\author{Dirk Grupe\altaffilmark{1},\email{grupe@astro.psu.edu}
David N. Burrows\altaffilmark{1}, 
Sandeep K. Patel\altaffilmark{2,3},
Chryssa Kouveliotou\altaffilmark{3},
Bing Zhang\altaffilmark{4},
Peter \meszaros\altaffilmark{1,5},
Ralph A.M. Wijers\altaffilmark{6},
Neil Gehrels\altaffilmark{7}
}

\altaffiltext{1}{Department of Astronomy and Astrophysics, Pennsylvania State
University, 525 Davey Lab, University Park, PA 16802} 
\altaffiltext{2}{Universities Space Research Association, 10211 Wincopin Circle, Suite 500
Columbia, MD 21044-3432}
\altaffiltext{3}{NASA/Marshall Space Flight Center, National Space Science
Technology Center, VP-62, 320 Sparkman Dr., Huntsville, AL 35805}
\altaffiltext{4}{Department of Physics, University of Nevada, Las Vegas, 
NV 89154}
\altaffiltext{5}{Department of Physics, Pennsylvania State University,
University Park, PA 16802}
\altaffiltext{6}{Astronomical Institute 'Anton Pannekoek', University of
Amsterdam, Kruislaan 403, NL-1098 SJ Amsterdam, The Netherlands}
\altaffiltext{7}{NASA Goddard Space Flight Center, Green Belt, MD 20771}




\begin{abstract}
We report the results of the \chandra\ observations of the \swift-discovered
short Gamma-Ray Burst
GRB\,050724. \chandra\ observed this burst twice, about two days after the burst
and a second time three weeks later. 
The first \chandra\ pointing 
occurred at the end of a strong late-time flare. About 150 photons were detected 
during this 49.3~ks observation  in the 0.4-10.0 keV range. 
The spectral fit is in good agreement
with spectral analysis of earlier \swift~XRT data.
In the second \chandra\ pointing the afterglow was clearly detected with 8 
background-subtracted photons in
44.6~ks.  
From the combined \swift~XRT and \chandra-ACIS-S light curve we find
significant flaring superposed on an underlying power-law decay slope of
$\alpha$=0.98$^{+0.11}_{-0.09}$.
There is no evidence for
a break between about 1~ks after the burst and the last \chandra\ pointing about
three weeks after the burst.  The non-detection of a jet break places a
lower limit of 25$^{\circ}$ on the jet opening angle, indicating that
the outflow is less strongly collimated than most previously-reported 
long GRBs. This implies that the beaming corrected energy of
GRB\,050724 is at least $4\times 10^{49}$ ergs.
\end{abstract}


\section{Introduction}

Gamma-Ray Bursts (GRBs) are separated into two classes 
\citep{chryssa93, paciesas99}: long,
soft bursts ($t_{90}>~3$\,s, where $t_{90}$ is the time interval 
within which 90\% of the flux arrives), 
which are associated with the formation of a
black hole during massive star core
collapse \citep{woosley93}; and short, hard bursts ($t_{90}<~3$\,s), 
thought to be the result of mergers of compact
objects \citep[e.g ][]{eichler89, paczynski91, gehrels05, fox05,
barthelmy05b, bloom06, berger05}.
Both types of GRBs are thought to produce a highly relativistic
fireball that results in the prompt $\gamma$-ray emission through
internal shocks in the outflow, and subsequently causes a
broad-band afterglow
when it shocks the surrounding medium \citep{meszaros97, zhang04}.
Much progress has been made since 1997 on the nature of long GRBs,
primarily through the study of their radio, optical, and X-ray
afterglows, but no afterglows of short GRBs had been detected until
2005, so our understanding of short GRB production mechanisms and
environments is still in its infancy.
The discoveries by \swift\ and HETE of the afterglows and counterparts
 for short GRB's 050509B \citep{gehrels05, bloom06}, 
 050709 \citep{villasenor05, fox05, hjorth05}, and 050724 \citep{barthelmy05b, berger05} 
  was a tremendous breakthrough in the field, 
on par with
the  discovery of 
the first  GRB afterglow by Beppo-SAX \citep[GRB 970228;][]{costa97}.
The X-ray light curve of GRB 050724
displays several flares including a late time flare
starting at  about 20\,ks after the detection of the burst and lasting for about a
day \citep{campana05b}. 
The early X-ray observations of GRB\,050724
by \swift\ also revealed the first \swift-discovered dust-scattered halo
\citep{romano05, vaughan05}, which was only the second time that an X-ray halo
was seen around a GRB afterglow. We report here on 
combined {\it Chandra} and {\it Swift}~XRT observations of the X-ray
afterglow of GRB\,050724, the first high-quality X-ray afterglow of a
short burst, 
and their implications for beaming in this object.

The paper is organized as follows: In \S\ref{observe} we describe the
observations and the data reduction.  In \S\ref{results} we
present the data analysis.  The discussion of our results is given in
\S\ref{discuss}. 
Throughout
the paper decay and energy spectral indices $\alpha$ and $\beta$ 
are defined by $F_{\nu}(t,\nu)\propto
(t-t_0)^{-\alpha}\nu^{-\beta}$, with $t_0$ the trigger time of the burst. 
Luminosities are calculated assuming a $\Lambda$CDM
cosmology with $\Omega_{\rm M}$=0.27, $\Omega_{\Lambda}$=0.73 and a Hubble
constant of $H_0$=71 km s$^{-1}$ Mpc$^{-1}$ using the luminosity distances
given by \citet{hogg99}. All errors are 1$\sigma$ unless stated otherwise. 
 
\section{\label{observe} Observations and data reduction}

The Burst Alert Telescope \citep[BAT, ][]{barthelmy05a} onboard the
\swift~Gamma-Ray-Burst-Explorer Mission \citep{gehrels04} triggered on
GRB\,050724 at 12:34:09 UT on 2005 July 24  \citep{covino05}. The burst had 
$T_{90}$=3.0\plm1.0\,s, but most of the flux was released in a hard spike with a
duration of 0.25\,s \citep{barthelmy05b}.
Therefore it was considered to be a short GRB. \swift's X-ray Telescope
\citep[XRT, ][]{burrows05} began observing the afterglow 74\,s after the trigger.
GRB\,050724 was not detected by \swift's
UV-Optical Telescope \citep[UVOT, ][]{roming05} and only a 3$\sigma$ limit of
V$>$18.84 was reported by \citet{chester05}. 
Spectroscopic redshifts were reported by \citet{prochaska05} (z=0.258\plm0.002)
and  \citet{berger05} (z=0.257\plm0.001), who also report on the radio and NIR
observations of GRB\,050724.

The \chandra\ X-ray Observatory performed two Target of Opportunity
observations of GRB\,050724.  The first occurred two days after
the burst at 2005 July 26 20:10 - 2005 July 27 10:45 (UT) for a total of
49.3\,ks.  The second observation was about three weeks after the burst for a
 total of 44.6 ks at
2005 August 14 20:29 - 2005 August 15 10:15. 
All observations were performed in
Very-Faint mode with the standard 3.2s read-out time on the
on-axis position on the back-illuminated ACIS-S3 CCD.
We reprocess all Chandra event data using CIAO $\sl acis_process_events$ and 
applied VF mode cleaning in order to reduce the ACIS particle background. 
Source photons were collected in a circular region with r=$1.5''$ and
r=$1.0''$ for the first and second \chandra\ observations, respectively.  The
background photons were selected from the primary event file in a circle 
with r=$15''$ and r=$10''$, respectively. The data reduction was performed
using the \chandra\ analysis software CIAO version 3.3. The X-ray spectrum from
the first \chandra\ observation was extracted with the CIAO tool 
{\it dmextract} and was
analyzed with {\it XSPEC} version 12.2.1 \citep{arnaud96}.
The calibration database was CALDB version
3.2.0. The response matrix and the auxiliary response file were created by the
CIAO tools {\it mkrmf} and {\it mkarf}. The spectrum was rebinned 
using the FTOOL {\it grppha} version 3.0.0 to have at least 15 photons per bin.

The \swift~XRT observed GRB\,050724 in the Windowed Timing (WT) and Photon
Counting (PC) observing modes \citep{hill04}. These X-ray observations have been
discussed in detail by \citet{campana05b} and in this paper we only concentrate
on the PC mode data of the later time flare (T$>$20 ks after the burst)
in order to compare the XRT data with our \chandra\ data.
The XRT data were reduced by the {\it xrtpipeline} task version 0.9.9. 
Source photons were selected by {\it XSELECT} version 2.3
in a circular region with a radius
of r=47$''$ and the background photons in a circular region close by with a
radius r=96$''$. For the spectral data only events with grades 0-12 were
selected with {\it XSELECT}.
The spectral data were re-binned by using 
{\it grppha} having 20 photons per bin. The spectra were 
analyzed with
{\it XSPEC} version 12.2.1. The auxiliary response files were created by {\it
xrtmkarf} and the standard response matrix swxpc0to12\_20010101v008.rmf
 was used.

Background-subtracted X-ray flux
light curves in the 0.3-10.0 keV energy range of the \chandra\ and
\swift\ observations were constructed using the ESO Munich
Image Data Analysis Software {\it MIDAS} (version 04Sep).
 The count rates were converted into unabsorbed
 flux units using energy conversion factors
(ECF) which were determined by calculating  the count rates and the
unabsorbed fluxes in the 0.3-10.0 keV energy band using XSPEC
as described in \citet{nousek05}. The ECFs are 
 8.3$\times 10^{-11}$ ergs s$^{-1}$ cm$^{-2}$ counts$^{-1}$ 
 for the \swift\ observation 
1.6$\times 10^{-11}$ ergs s$^{-1}$ cm$^{-2}$ counts$^{-1}$ for the \chandra\
observations.

\section{\label{results} Data Analysis}

\subsection{Position}
The first \chandra\ observation found a single source located within
the \swift~XRT error circle at
RA (J2000) = $16^{\rm h} 24^{\rm m} 44.^{\rm s}36$,
Dec (J2000) = $-27^{\circ} 32' 27 \farcs 5$,
with an error radius of $0 \farcs 5$ \citep{burrows05d}. 
This places the X-ray afterglow $1 \farcs 0$ south of the center 
of an elliptical galaxy \citep{barthelmy05b, berger05}, coincident
with the optical afterglow position reported by \citet{berger05} and
with the VLA radio position given by \citet{berger05} and
\citet{soderberg05b}. 

\subsection{Light curve}

\begin{figure*}
\epsscale{0.75}
\plottwo{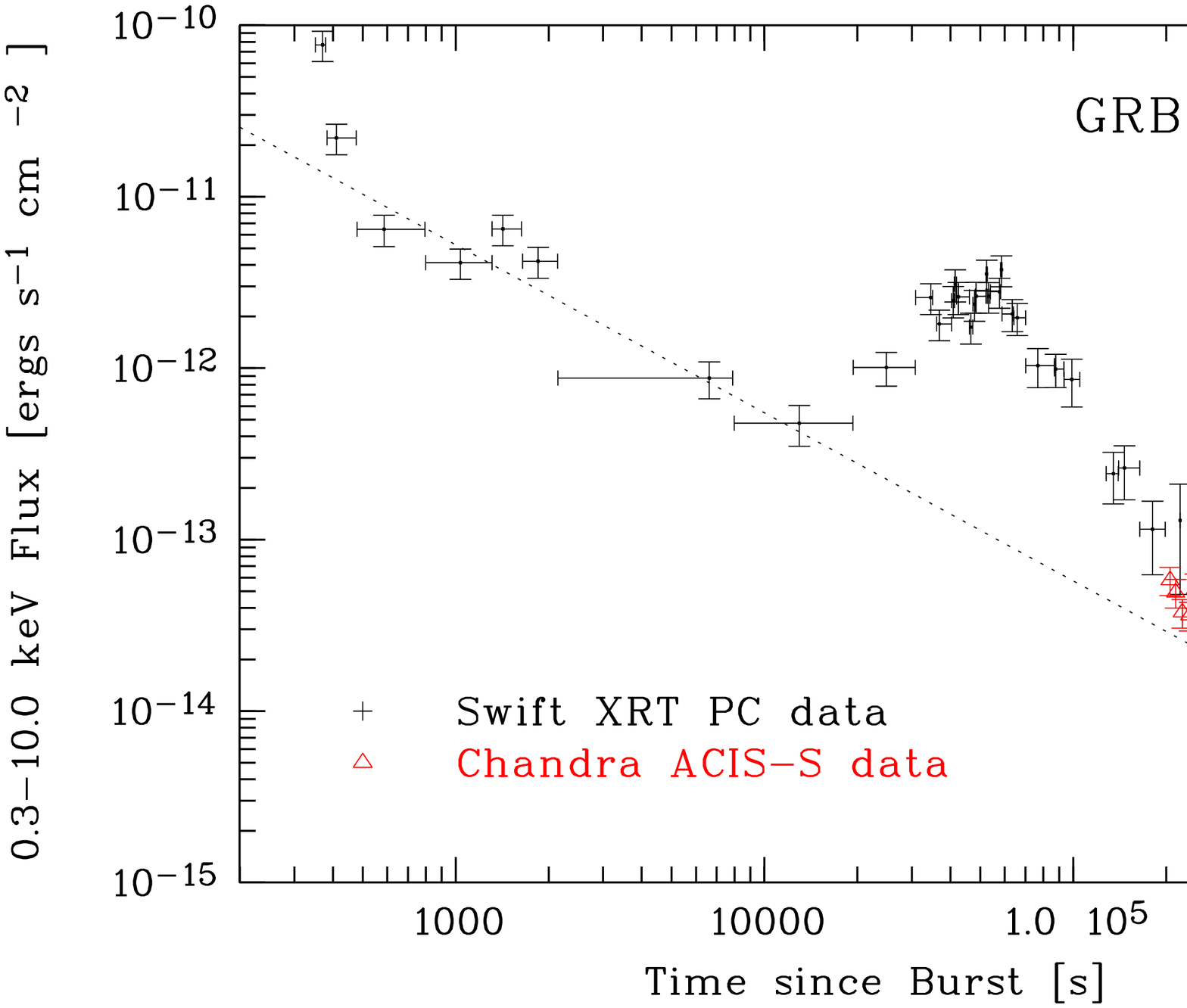}{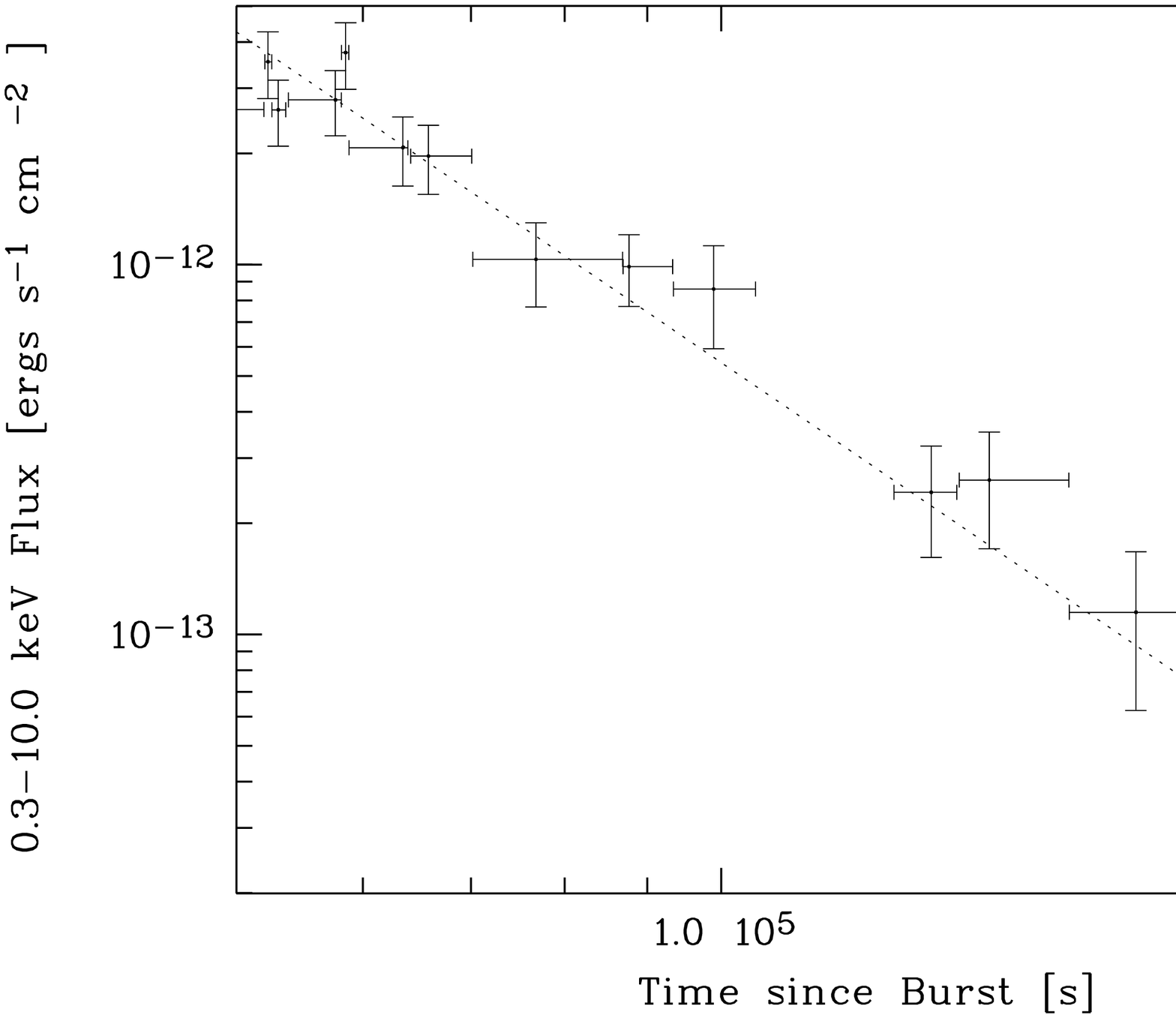}
\caption{\label{grb050724_lc} \swift~XRT PC mode and \chandra~ACIS-S light
curve of the afterglow of GRB\,050724. The light curve shows the observed 0.3-1.0
keV unabsorbed flux. The right panel displays the \swift~XRT and
\chandra~ACIS-S light curves of the decline of the 
late time flare. The dotted lines display
the decay slopes of the afterglow (left, $\alpha$=0.98$^{+0.11}_{-0.09}$), 
and the flare decay (right, $\alpha=2.98^{+0.16}_{-0.13}$).
}
\end{figure*}

The left panel of 
Figure~\ref{grb050724_lc} displays the 0.3-10\,keV unabsorbed flux light curve
of the \swift~XRT PC mode and \chandra~ACIS-S observations. Table\,\ref{lc_all}
lists the times and fluxes of the \swift\ and \chandra\ observations.
Following the early very steep decay (between 100 and 300\,s;
see \citet{campana05b}), the light curve is well fit by a single
underlying power law decay slope of 
$\alpha$=0.98$^{+0.11}_{-0.09}$
(indicated by the dotted line in Figure\,\ref{grb050724_lc}), with
one small and one very large flare superposed. The slope was determined from a
linear regression fit to the four underlying afterglow data points in the 
\swift~XRT
light curve plus the final \swift~XRT data point. Note, that the last XRT data
point is just a marginal detection at the 2$\sigma$ level containing 
a total observing time of 47.1 ks. 
The afterglow decay slope $\alpha$=0.98
deviates from the slope $\alpha$=0.60\plm0.20 given in \citet{campana05b}. 
We analyzed our afterglow light curves by using different fitting
routines within XSPEC, IDL, and MIDAS and always found a decay slope
$\alpha\approx$1.0.  Most likely the difference between our 
decay slope and that found  
by \citet{campana05b} is due to different binning 
methods and/or treatment of the various flares that complicate modeling the 
underlying afterglow.

The largest
flare occurs about 50ks after the burst. This flare has a
fluence of 7.7$\times 10^{-8}$ ergs cm$^{-2}$ 
\citep{campana05b}.
The first \chandra\ observation from 2005-July-26/27 occurred at the end of this
late-time flare. 
While the afterglow was almost undetectable with the
\swift~XRT, \chandra\ was able to detect 150 source photons in the 49.3 ks
observation in the 0.4-10.0 keV band.
A
combined XRT and ACIS-S light curve of the decaying flare as displayed in the
right panel of Figure\,\ref{grb050724_lc}
results in a decay
slope $\alpha$=2.98$^{+01.6}_{-0.13}$ ($\chi^2/\nu=9.8/14$).

During the second \chandra\ observation three weeks after the burst
8 background-subtracted source photons were detected in 44.6 ks. 
As shown in the left panel of Figure~\ref{grb050724_lc} 
this last data point is consistent with a flat decay slope
$\alpha$=0.98, representing the underlying afterglow due to the
forward external shock in the ambient medium.
There is no sign of a jet break in the light
curve up to the final \chandra\ point three weeks after the burst.

\subsection{Spectral Analysis}

\begin{figure*}
\epsscale{0.9}
\plottwo{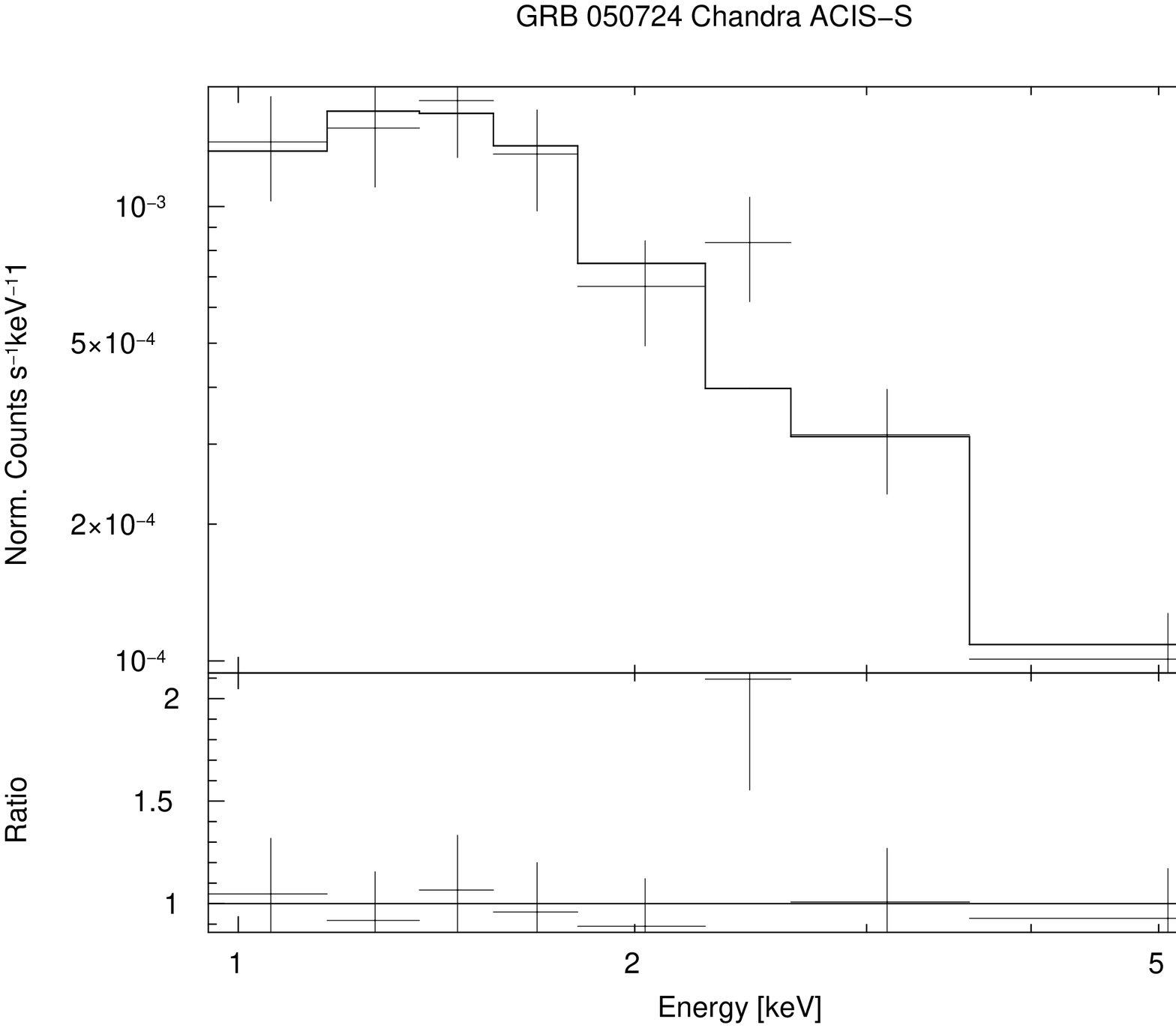}{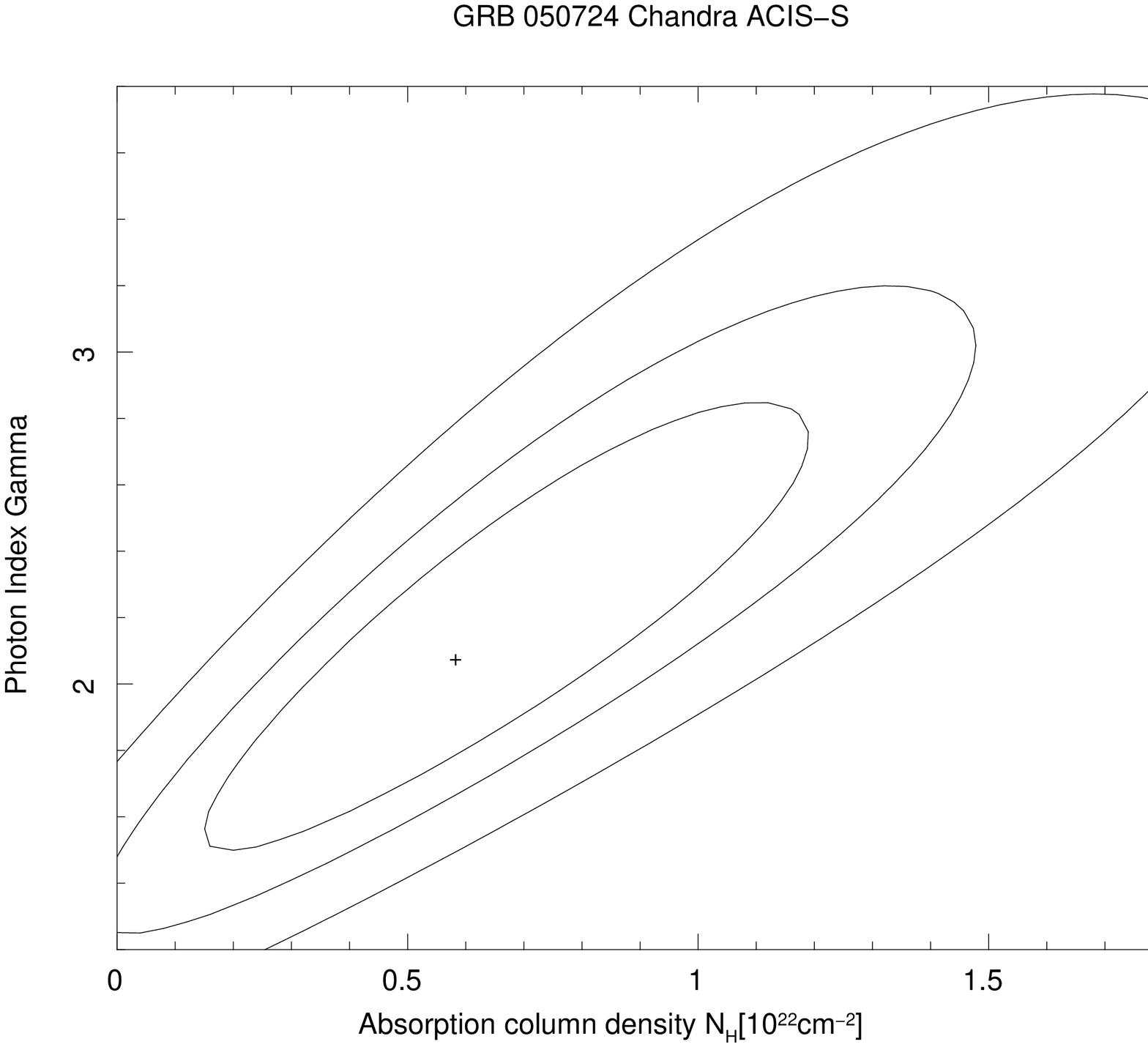}
\caption{\label{grb050724_acis_spec} \chandra~ACIS-S3 spectrum. The left panel
displays the X-ray spectrum with an absorbed power law model fitted to the data.
The right panel shows the contour plot between the absorber column density and
the photon index $\Gamma = \beta_{\rm X}$ + 1. The Galactic column
density as given by \citet{vaughan05} is $N_{\rm H} = 3.4-4.2 \times 10^{21}$
cm$^{-2}$. The contour levels are 1, 2, and 3 $\sigma$.
}
\end{figure*}

Figure~\ref{grb050724_acis_spec} displays the \chandra~ACIS-S spectrum of 
GRB\,050724. This spectrum was obtained during the decay of the late-time flare in the
light curve of GRB\,050724 (left panel of Figure~\ref{grb050724_lc}).
This spectrum can be
well-fitted by a single absorbed power law (using $wabs*powl$ in {\it XSPEC}). 
The results of the fits to the X-ray
data are listed in Table~\ref{spectral}. A free fit to the data results in an
absorption column density at z=0 of \nh=5.86$^{+6.32}_{-2.99} \times 10^{21}$
cm$^{-2}$,
which is significantly in excess of the Galactic value 
\nh=1.48$\times 10^{21}$ cm$^{-2}$ given by \citet{dic90}. However, as noticed
by \citet{vaughan05}, the line-of-sight of GRB\,050724 passes close to the
Ophiuchus molecular cloud and the Galactic column density is therefore
significantly higher than what is given by \citet{dic90}. \citet{vaughan05}
estimated the real value of the Galactic column density to be 3.4-4.2$\times
10^{21}$ cm$^{-2}$. Therefore the fit to the \chandra~ACIS-S data is consistent
with the Galactic value and no additional absorption is required. 
Figure~\ref{grb050724_acis_spec} displays the single power law fit to the 
\chandra~ACIS-S data.
The left panel of Figure~\ref{grb050724_acis_spec} shows the ACIS-S spectrum and
the right panel displays a contour plot of the absorption column density \nh~and
the photon spectral index $\Gamma$=\bx+1. We conclude that no
intrinsic absorber is required in the host galaxy. 
The X-ray energy
spectral slope as listed in Table~\ref{spectral}
is \bx=0.86$^{+0.34}_{-0.31}$, where the absorption column density
is fixed to the Galactic value with 
\nh=4.0$\times 10^{21}$cm$^{-2}$ as given
by \citet{vaughan05}. 

\begin{figure*}
\epsscale{0.9}
\plottwo{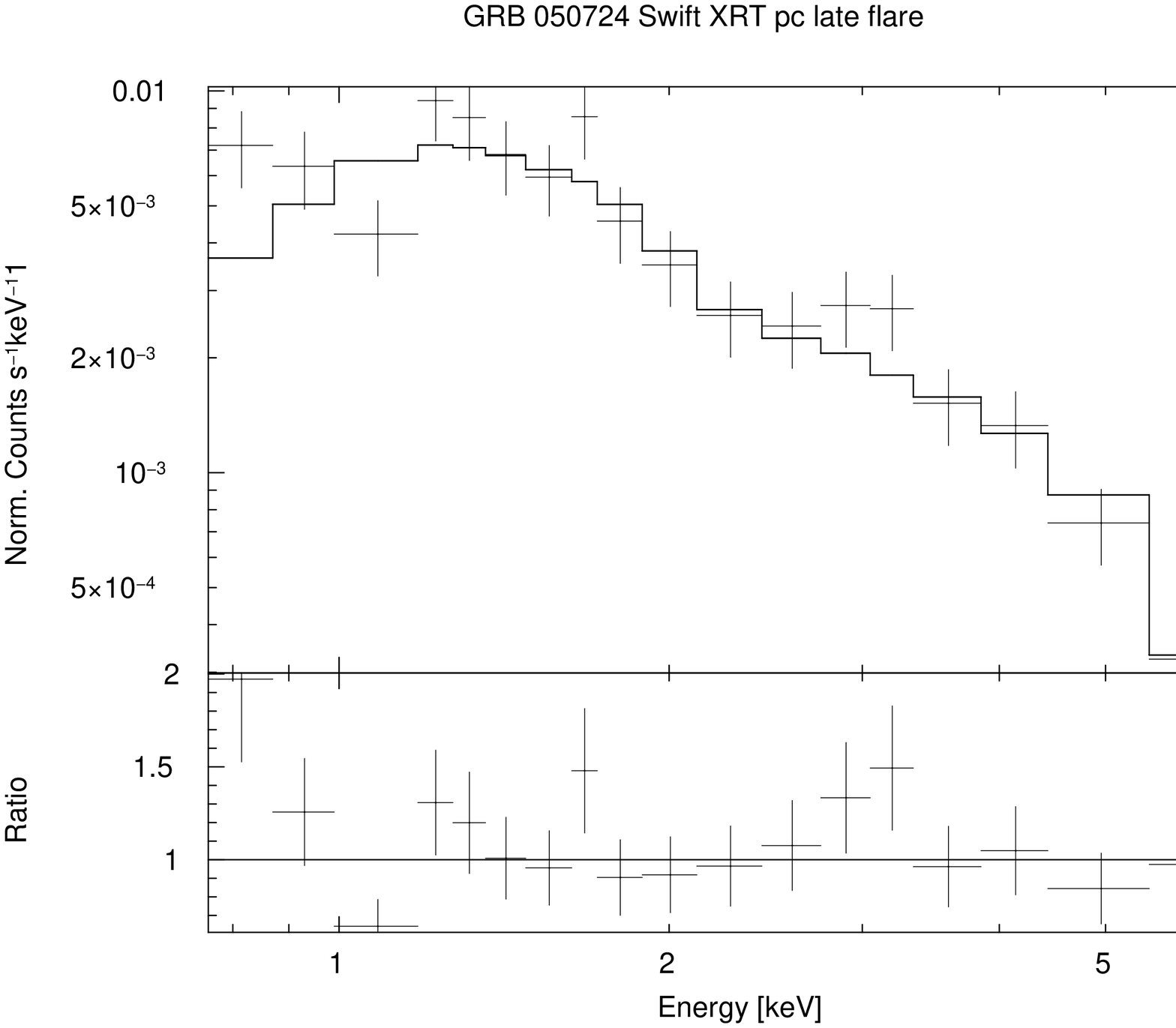}{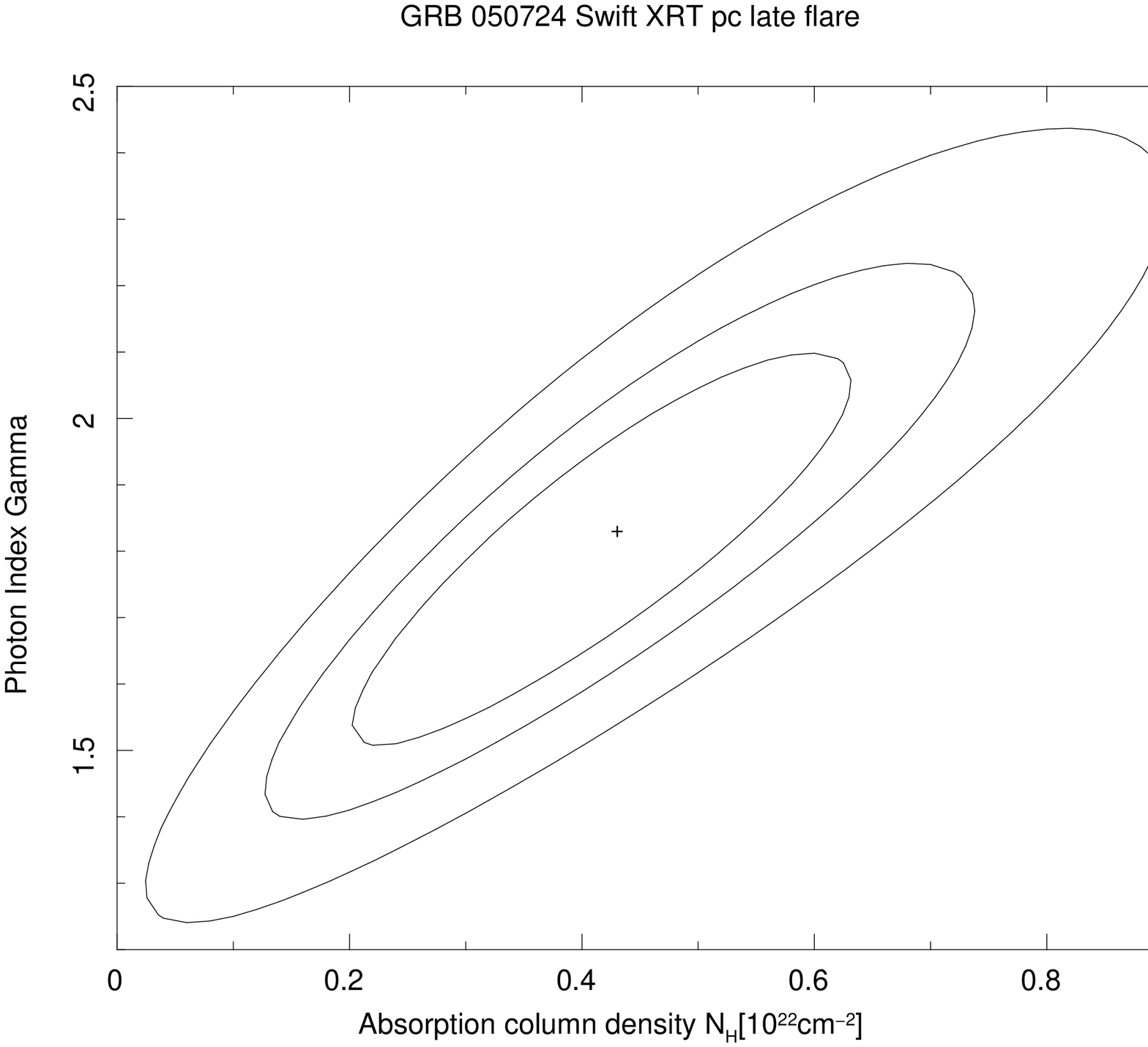}
\caption{\label{grb050724_xrt_spec} \swift~XRT PC mode spectrum. The left panel
displays the X-ray spectrum with an absorbed power law model fitted to the data.
The right panel shows the contour plot between the absorber column density and
the photon index $\Gamma = \beta_{\rm X}$ + 1. The Galactic column
density as given by \citet{vaughan05} is $N_{\rm H} = 3.4-4.2 \times 10^{21}$
cm$^{-2}$. The contour levels are 1, 2, and 3 $\sigma$.
}
\end{figure*}

For comparison, we examine the \swift~XRT 
spectrum of the late-time flare of GRB\,050724, shown in
Figure~\ref{grb050724_xrt_spec}.  As listed in Table~\ref{spectral},
our analysis of the XRT data of the late time flare (T$>$20 ks after the burst)
results in an energy spectral index \bx=0.85$^{+0.28}_{-0.36}$ when the
absorption column density is left as a free parameter. The absorption column
density \nh=4.44$^{+2.13}_{-2.52}\times10^{21}$cm$^{-2}$ also agrees 
with the Galactic absorption column density as given by \citet{vaughan05} and
does not require any additional absorption. This result
agrees well with the analysis reported by \citet{campana05b}.
The right panel of Figure~\ref{grb050724_xrt_spec} displays the contour plot
between the absorption column density \nh~ and the photon index
$\Gamma$. 

Table\,\ref{spectral} also lists the results from a joint fit to the
\chandra~ACIS-S and \swift~XRT spectra. This joint fit yields similar
results to those obtained from the separate spectral fits.

The \chandra\ spectral analysis from the fading tail of this flare is in
excellent agreement with the spectral fit for the flare as a whole, confirming
the conclusion of \citet{campana05b} that
there is no evidence for spectral
evolution in this afterglow.

\section{\label{discuss} Discussion}

The main result of our late time \chandra\ observation is that there is no jet
break observed in the X-ray afterglow of GRB\,050724.  
Our last Chandra data point connects to the lower
envelope of the XRT data points before $\sim 20$~ks as a single power law decay with
$\alpha$=0.98$^{+0.11}_{-0.09}$.
We interpret this slope as the afterglow from the forward shock,
extending from $\sim 0.5$~ks after the burst until after the
last \chandra\ observation 3 weeks later, with flaring superposed on
this steady decay.  Following the slow cooling ISM case listed in 
Table 1 in \citet{zhang04} we find that the slope of the electron spectrum is
$p=2.70$ and that the observed
emission frequency $\nu$ is $\nu_{\rm m} <
\nu < \nu_{\rm c}$ with $\nu_{\rm m}$ the injection frequency and 
$\nu_{\rm c}$ the
cooling frequency \citep[See ][ for details.]{zhang04}.

The non-detection of a jet break up until 22 days after the burst 
places constraints
on the jet opening angle. Using the relations given by \citet{sari99} and 
\citet{frail01}, we can estimate the jet opening
angle as: 
\begin{equation}
\Theta > 25^{\circ}\left(\frac{n}{0.1}\right)^{1/8}\left(\frac{E_{\gamma, \rm iso}}{4\times 10^{50}}\right)^{-1/8} 
\end{equation}
using an ambient medium density
of $n=0.1$ cm$^{-3}$, an isotropic equivalent energy of 4
$\times 10^{50}$ ergs s$^{-1}$ and an isotropic-equivalent kinetic energy of
$E_{\rm k, iso}$ = 1.5$\times 10^{51}$ ergs, and an efficiency of the fireball
$\eta = \frac{E_{\rm K, iso}}{E_{\rm \gamma, iso}}$=0.2,
as given by \citet{berger05}. 
If the density is as high as $10^3$ cm$^{-3}$ \citep{panaitescu05},
the inferred jet angle must be larger than $79^{\circ}$.

This result differs from the conclusions of
\citet{panaitescu05} and \citet{berger05}, 
who estimated an opening angle of 8$^{\circ} -
12^{\circ}$ based on the radio and NIR
observations of the afterglow of GRB\,050724. The
radio data seem to suggest an early jet break about 1--2 days after the burst, based
on the steep radio decay slope $\alpha_{\rm radio}$=2.0, 
but the paucity of both the optical and radio
observations and the possibility of strong Galactic interstellar
scintillation in the 8.5~GHz radio band make this conclusion uncertain
\citep{panaitescu05}.
In addition, we point out that both the optical and radio data
occur on the declining side of
the large X-ray flare.
If this flare is caused by central engine activity, the optical and radio
behavior may be related to this flare rather than to a jet break.  
In any case, since a jet break is expected to be achromatic, the \chandra\ data
strongly rule out a jet break at one day post-burst. The energy of the prompt
emission is $3\times10^{50}$ ergs \citep{barthelmy05b} while the energy in the
late-time flare is $2\times10^{49}$ ergs \citep{campana05b} which is about 7\%
of the energy of the prompt emission. This result supports  central engine
models which predict a reduced activity at later times \citep[e.g. ][]{zhang04}.

Until recently, no jet break has ever been convincingly found in a short GRB
afterglow. \citet{fox05} discussed the possibility of a jet break in the
HETE-2-discovered short burst GRB 050709. However, their conclusions
are based
on few data points that poorly constrain jet parameters \citep{panaitescu05} and may be open to other interpretations,
particularly given the complex nature of X-ray afterglows, even for
short GRBs (e.g. GRB\,050724). The only
convincing case of a jet break found in a short GRB afterglow is GRB 051221A
\citep[][ Paper II]{burrows06},
 based on detailed, well-sampled \swift\ and \chandra\ observations.

The large lower limit of the jet angle inferred from our data suggests
that the jet of this particular short GRB is much less
collimated than typical long GRBs reported in previous works
\citep{frail01,bloom03,soderberg06}, although we note that \swift\ afterglows in
general may not support those early results (Sato et al. 2006, in
preparation; Willingale et al. 2006, in preparation).
The elliptical host galaxy of GRB\,050724
suggests that the progenitor is not a collapsar and is likely a compact 
star merger
(NS-NS or NS-BH). The wide jet angle inferred from the data is
consistent with such a progenitor scenario 
\citep[e.g. ][]{meszaros99},
since there is no extended massive stellar envelope as in long
GRBs that serves to naturally collimate the outflow \citep[e.g. ][]{wzhang04}.
 Numerical simulations of mergers \citep[e.g. ][]{aloy05}
suggest a varying degree of short GRB collimation angles,
which are in general wider than those of long GRBs. A wide angle also
potentially explains why short GRBs are less
bright \citep[e.g. ][]{meszaros99}. 
The total
energy budget (collimation corrected) of short GRBs then should
not be orders of magnitudes lower than that of long GRBs. 
Swift observations show
that short GRBs typically have a very low isotropic gamma-ray energy
given their very low redshifts \citep[e.g. ][]{gehrels05, barthelmy05b, fox05}.
However, if short GRBs typically have wide jets, this would reduce
the large gap in the total energy budget of short
and long GRBs implied by the isotropic energies, which differ by several orders
of magnitude \citep{frail01} under the assumption of similar jet angles.
Applying the beaming correction of $(1- cos \Theta)$ as given by 
\citet{sari99} and \citet{rhoads99} and using the isotropic energy given by
\citet{berger05} of $E_{\gamma, \rm iso} = 4\times10^{50}$ ergs, the beaming
corrected energy of GRB\,050724 is $4\times10^{49}$ to $4\times10^{50}$
ergs, depending on the beaming angle. 
This energy is
comparable with the modest fireball energies predicted from neutron star -
neutron star mergers based on neutrino - anti-neutrino annihilation as
given by, e.g., 
\citet{popham99}.

The long measurement baseline for the afterglow slope enabled by
the late \chandra\ observation also guides the interpretation of
the late flare, making it clear that the 
rapid decline in the X-ray light curve after 60~ks is the decay of a
flare and not a jet break.  In the \swift\ era,
X-ray flares have generally been interpreted as due to late
central engine activity \citep{burrows05c, bing05, falcone06, romano06}.
\citet{bing05} pointed out that
the steep decay ($\alpha \sim 2.70$) after this flare may be an indication
of the curvature effect of a late internal dissipation activity. \citet{liang06}
have tested such a model by searching for the required
time zero point of this flare to satisfy the curvature effect
interpretation and found that it is right at the rising part of the
flare. Our late Chandra data point suggests that this flare is an
independent component that is superimposed on the otherwise power-law
decaying afterglow component. This flare may then originate
from a different emission site, adding credence to the late central
engine activity interpretation. 

This is a particularly long and bright flare for such late times, and it is
surprising to find it in a short GRB afterglow, since models of NS-NS
mergers have typically suggested rapid creation of a black hole with
little material left over to feed the central engine at such late
times.
These observations suggest that the
central engine of GRB\,050724 may have reactivated at around a day
after the burst, which
calls for central engine models that can extend the short GRB central
engine to such a late time. 
Possible scenarios include fragmentation
of a neutron star by a black hole in a NS-BH merger \citep{Faber06}, 
fragmentation of the accretion disk \citep{perna06}, 
a magnetic-barrier-modulated accretion flow \citep{proga06}, or the magnetic
activity of a post-merger massive neutron star \citep{dai06}.
Future extensive afterglow observations and more detailed theoretical
modeling will be required to further understand the progenitor 
and the central engine of short GRBs.

Our observations of GRB\,050724 have demonstrated the importance of late time
\chandra\ observations of GRB afterglows. 
Our last \chandra\ observation from 2005 August provides crucial information
about the late time behavior of the afterglow that no other X-ray observatory is 
able to measure at these low flux levels. Further late time X-ray
observations of GRB afterglows will provide an improved understanding of the
characteristics of jets of both long and short GRBs.

\acknowledgments

We would like to thank the anonymous referee for varous comments and suggestions to
improve the paper, and Sergio Campana for discussions on the XRT light
curve.
This research has made use of data obtained through the High Energy 
Astrophysics Science Archive Research Center Online Service, provided by the 
NASA/Goddard Space Flight Center.
This research was supported by NASA contract NAS5-00136 and SAO
grant G05-6076X BASIC.



\begin{deluxetable}{lrrcc}
\tablecaption{\swift~XRT and \chandra\ light curve data of GRB\,050724 
\label{lc_all}}
\tablewidth{0pt}
\tablehead{
\colhead{Observatory} & \colhead{Time after burst\tablenotemark{1}} &
\colhead{$T_{\rm obs}$\tablenotemark{1}} & 
\colhead{Flux\tablenotemark{2}}  &
\colhead{Flux error\tablenotemark{2}} 
} 
\startdata
\swift &     348 &    15 &  1.40$\times 10^{-10}$ &  0.28$\times 10^{-10}$ \\
       &     370 &    28 &  7.68$\times 10^{-11}$ &  1.53$\times 10^{-11}$ \\
       &     410 &    95 &  2.20$\times 10^{-11}$ &  0.45$\times 10^{-11}$ \\
       &     586 &   318 &  6.45$\times 10^{-12}$ &  1.33$\times 10^{-12}$ \\
       &    1035 &   514 &  4.12$\times 10^{-12}$ &  0.82$\times 10^{-12}$ \\
       &    1421 &   323 &  6.48$\times 10^{-12}$ &  1.31$\times 10^{-12}$ \\
       &    1847 &   493 &  4.20$\times 10^{-12}$ &  0.86$\times 10^{-12}$ \\
       &    6625 &  2033 &  8.75$\times 10^{-13}$ &  2.12$\times 10^{-13}$ \\
       &   12960 &  3415 &  4.77$\times 10^{-13}$ &  1.27$\times 10^{-13}$ \\
       &   24813 &  1910 &  1.00$\times 10^{-12}$ &  0.22$\times 10^{-12}$ \\
       &   34556 &   805 &  2.58$\times 10^{-12}$ &  0.53$\times 10^{-12}$ \\
       &   36840 &  1158 &  1.81$\times 10^{-12}$ &  0.37$\times 10^{-12}$ \\
       &   40898 &   830 &  2.47$\times 10^{-12}$ &  0.51$\times 10^{-12}$ \\
       &   41485 &   652 &  3.08$\times 10^{-12}$ &  0.65$\times 10^{-12}$ \\
       &   42476 &   772 &  2.60$\times 10^{-12}$ &  0.55$\times 10^{-12}$ \\ 
       &   46608 &  1196 &  1.73$\times 10^{-12}$ &  0.36$\times 10^{-12}$ \\
       &   47881 &   880 &  2.35$\times 10^{-12}$ &  0.48$\times 10^{-12}$ \\
       &   48481 &   790 &  2.62$\times 10^{-12}$ &  0.54$\times 10^{-12}$ \\
       &   52419 &   587 &  3.53$\times 10^{-12}$ &  0.72$\times 10^{-12}$ \\
       &   53191 &   800 &  2.62$\times 10^{-12}$ &  0.53$\times 10^{-12}$ \\
       &   57706 &   759 &  2.79$\times 10^{-12}$ &  0.56$\times 10^{-12}$ \\
       &   58551 &   554 &  3.74$\times 10^{-12}$ &  0.77$\times 10^{-12}$ \\
       &   63527 &   972 &  2.07$\times 10^{-12}$ &  0.44$\times 10^{-12}$ \\
       &   65890 &  1023 &  1.96$\times 10^{-12}$ &  0.42$\times 10^{-12}$ \\
       &   76802 &  1639 &  1.03$\times 10^{-12}$ &  0.26$\times 10^{-12}$ \\
       &   87692 &  1973 &  9.88$\times 10^{-13}$ &  2.16$\times 10^{-13}$ \\
       &   98926 &  1208 &  8.59$\times 10^{-13}$ &  2.66$\times 10^{-13}$ \\
       &  134892 &  4438 &  2.43$\times 10^{-13}$ &  0.81$\times 10^{-13}$ \\
       &  146538 &  3952 &  2.62$\times 10^{-13}$ &  0.91$\times 10^{-13}$ \\
       &  180599 &  6980 &  1.15$\times 10^{-13}$ &  0.53$\times 10^{-13}$ \\
       &  221928 &  2121 &  1.29$\times 10^{-13}$ &  0.81$\times 10^{-13}$ \\
       &  560000 & 47109 &  1.30$\times 10^{-14}$ &  0.75$\times 10^{-14}$ \\
\chandra &205920 &  8270 &  5.80$\times 10^{-14}$ &  1.08$\times 10^{-14}$ \\
         &214612 &  9610 &  4.93$\times 10^{-14}$ &  0.93$\times 10^{-14}$ \\
         &225551 & 12546 &  3.76$\times 10^{-14}$ &  0.72$\times 10^{-14}$ \\
         &238353 & 13065 &  3.62$\times 10^{-14}$ &  0.69$\times 10^{-14}$ \\
         &248605 &  6465 &  5.14$\times 10^{-14}$ &  1.16$\times 10^{-14}$ \\
       & 1867280 & 43218 &  3.12$\times 10^{-15}$ &  1.20$\times 10^{-15}$

\enddata

\tablenotetext{1}{In units of s.}
\tablenotetext{2}{In units of ergs s$^{-1}$ cm$^{-2}$.}

\end{deluxetable}

\begin{deluxetable}{lccc}
\tablecaption{Power law fits to the X-ray spectrum of GRB\,050724 
\label{spectral}}
\tablewidth{0pt}
\tablehead{
\colhead{Observation} & \colhead{\nh\tablenotemark{1}} &
\colhead{$\beta_{\rm X}$} & 
\colhead{$\chi^2/\nu$} 
} 
\startdata
\chandra~ACIS-S & 5.86$^{+6.32}_{-2.98}$ & 1.08$^{+0.82}_{-0.42}$ & 4.6/5 \\
               & 4.00 (fix)             & 0.86$^{+0.34}_{-0.31}$ & 5.3/6 \\
\swift~XRT      & 4.44$^{+2.13}_{-2.52}$ & 0.85$^{+0.28}_{-0.36}$ & 20/15 \\
               & 4.00 (fix)             & 0.79$^{+0.16}_{-0.15}$ & 20/16 \\
ACIS-S + XRT    & 4.41$^{+2.49}_{-1.49}$ & 0.85$^{+0.29}_{-0.14}$ & 26/21 \\
               & 4.00 (fix)             & 0.81$^{+0.15}_{-0.14}$ & 26/22	      
\enddata

\tablenotetext{1}{Galactic absorption column density in units of $10^{21}$
cm$^{-2}$.
For the Galactic value we use  $4.0 \times 10^{21}$ cm$^{-2}$  
as given by \citet{vaughan05}
}

\end{deluxetable}

\end{document}